\documentclass[useAMS,usenatbib]{mn2e}
\usepackage{graphicx}
\usepackage{graphics}
\usepackage{epstopdf}
\usepackage{float}
\usepackage{bm}
\usepackage{epsf}
\usepackage{url}

\newcommand{\apj}{{ApJ}}
\newcommand{\apjl}{{ApJL}}
\newcommand{\apjs}{{ApJS}}
\newcommand{\mnras}{{MNRAS}}
\newcommand{\nat}{{Nature}}
\newcommand{\aap}{{A\&A}}

\newcommand{\jcap}{{JCAP}}
\newcommand{\prd}{{Phys. Rev. D}}

\title[The MIP Ensemble Simulation: Local Ensemble Statistics in the Cosmic Web]{The MIP Ensemble Simulation: Local Ensemble Statistics in the Cosmic Web}

\author[Aragon-Calvo M.A.]{M.A. Aragon-Calvo$^{1,2}$\thanks{E-mail:maragon@ucr.edu} \\
$^{1}$University of California, Riverside,  900 University Ave, Riverside, CA, 92521, USA.\\
$^{2}$The Johns Hopkins University, 3400 Charles St., Baltimore, MD, 21218,  USA.\\}
\begin{document}

\date{Accepted}

\pagerange{\pageref{firstpage}--\pageref{lastpage}} \pubyear{2002}
\maketitle
\label{firstpage}

%========================================================================
% ABSTRACT
%========================================================================
\begin{abstract}

We present a new technique that allows us to compute ensemble statistics
on a local basis, directly relating halo properties to their local environment. This is achieved by the use 
of a correlated ensemble in which the Large Scale Structure is common to all realizations while having each an independent halo population.
The correlated ensemble can be stacked, effectively increasing the halo number density by an arbitrary factor, thus 
breaking the fundamental limit in the halo number density given by the haloe mass function.
This technique allows us to compute \textit{local ensemble statistics} of the matter/haloe distribution
at any position in the simulation box, while removing the intrinsic stochasticity in the halo formation process and
directly relating halo properties to their environment.

We introduce the \textit{Multum In Parvo} (MIP) correlated ensemble simulation consisting of 220 realizations 
on a 32 h$^{-1}$ Mpc box with $256^3$ particles each. This is equivalent in terms of effective volume and number of 
particles to a box of $\sim 193$ h$^{-1}$ Mpc of side with $\sim 1540^3$ particles containing 
$\sim 5\times 10^6$  haloes with a minimum mass of $3.25 \times 10^9$ h$^{-1}$ M$_{\odot}$.

The potential of the technique presented here is illustrated by computing the local ensemble statistics of the halo ellipticity 
and halo shape-LSS alignment. We show that, while there are general trends in the ellipticity and alignment of haloes with their LSS, 
there are also significant spatial variations which has important implications for observational studies of galaxy shape and alignment.

\end{abstract}

%==========================================================================
%--- Introduction
%==========================================================================
\section{Introduction}

N-body simulations are one of our most valuable tools in cosmology. They allow us to follow the evolution of
cosmic structures from the linear regime to the present time over a wide range of scales and masses.
Current state-of-the-art N-body simulations contain billions of particles and can resolve from large superclusters and clusters of galaxies down to subhaloes and  dwarf galaxies \citep{Springel05, Diemand07, Springel08}.

One area where N-body simulations have been particularly useful is the study of the formation and evolution
of haloes and their relation with their large-scale environment. Haloes are affected by their environment in many ways 
including local density, tidal field, matter accretion and mergers
(\citet{White84, Byrd90,  Cole93, Zabludoff98, Bosch02,  Gottlober01} among others).
In order to properly understand the role of local environment in the process of galaxy formation and evolution
it is crucial to be able to relate the properties of haloes to their surrounding LSS.
An important step in this direction is the use of constrained simulations used to interpret real observations 
of specific (and often complex) environments such as our local neighborhood.
Even with sophisticated models of the local environment there are still many difficulties 
as the ``problems" existing between observations and simulations show.
Examples of such discrepancies include the ``cold Hubble flow" 
problem \citep{Sandage72,Sandage86,Aragon11a}, the ``local velocity anomaly" \citep{Tully08}, the peculiar galaxy population in
the local sheet \citep{Peebles10} and, to some degree, the ``missing satellite" problem  \citep{Klypin99}. 

Current theoretical and numerical studies of the LSS-halo relation are mostly focused on global descriptors
such as the two-point correlation function and its generalization to $n-$points \citep{Peebles80, Szapudi98}, 
marked correlations \citep{Sheth05} and topology via the Minkowski functionals \citep{Mecke94,Schmalzing99}. 
These tools provide a quantitative description of the global clustering and scaling relations
of the distribution of matter and galaxies/haloes in the Universe. However, such measurements are intrinsically global
and are insensitive to local effects that may arise in the diversity of environments in 
the Cosmic Web.  Some important advances have been done in this respect with the introduction of LSS characterization algorithms
based on the local properties of the density field \citep[MMF][]{Aragon07b}) and derived implementations
\citep{Zhang09, Cautun13}, tidal field \citep{Hahn07}, shear tensor \citep{Libeskind12}, topology \citep{Sousbie08,Aragon10a}. 
Using such tools haloes can be assigned to characteristic environments in order to study the dependence of their properties. However, in these methods the halo sample is invariably integrated into global samples such as ``filament haloes" or ``void haloes", 
thus partiality losing the locality gained by the LSS characterization. 

%--------------------------------------
%
%--------------------------------------
\subsection{Overcoming finite halo sampling}

The problem of correlating halo properties with their particular environment can be then described (after proper LSS characterization) as a sampling problem. 
Local studies of haloes and their environment are ultimately limited by the halo mass function \citep{Press74} which
defines the mean number density of haloes in a given mass range. The halo mass function is an intrinsic property of the density field and the underlying cosmology. As such it does not depend on the mass resolution imposed on the simulation. 
This sets fundamental limits to the kind of halo analysis we can perform on a local basis. One could for instance measure 
statistics of haloes at a given position using a sampling window small enough to be sensitive to local variations. While this may work for 
limited cases with a high halo number density, in general one does not have enough haloes to obtain statistically significant results from a single-realization. This limitation is particularly important
in low-density regions such as voids and walls where the number density of haloes massive enough to host luminous galaxies is
of the order of one halo per several cubic megaparsecs. 

In this paper we present a new technique that solves the limitations imposed by the halo mass function and, in particular, the finite halo number density. The technique is based
on a correlated ensemble simulation where all semi-independent realizations share the same large-scale fluctuations.
Ensemble simulations are becoming an important tool to study the statistical properties of the matter/galaxy distribution
where a large sampling volume is required \citep{Meiksin99,Takahashi09,Smith09, Schneider11,Forero11, Orban12}. 
For instance, \citet{Takahashi09} ran 5000 independent realizations of a 1 h$^{-1}$ Gpc box with $256^3$ particles
each to study the Baryon Acoustic Oscillations \citep[see also][]{Schneider11}. At smaller physical scales, \citet{Forero11} 
used ensemble simulations from constrained random realizations of the local supercluster in order to identify structures of 
interest which were subsequently run at higher resolution.  An interesting and promising line of work related to the one presented here is the analysis of the local supercluster by \citet{Kitaura12} 
where they use Bayesian Networks Machine Learning to reconstruct the local volume and generate an ensemble of 100 constrained random realizations of the local supercluster to study local statistics of the environment around the Milky Way.

%--------------------------------------
%
%--------------------------------------
\subsection{Applications of the MIP}

The technique and simulation presented in this paper has been already applied to study the bias deep inside voids with practically no stochasticity by taking advantage of the high local number density of haloes even at the centres of voids. Since bias is a local effect we can not simply average over a large number of voids since we would have to compute the density of haloes {\it per void} which in standard simulations is dominated by shot noise \citep{Neyrinck2013}. 
Using standard techniques one could compute the bias of haloes in voids by stacking a large number of voids, but this assumes that all voids are similar or can be fully described by a few properties such as radius, mean density or shape. This assumption is not necessarily correct as voids with similar properties may be surrounded by very different large scale structures that may affect the void's haloes population. This also ignores the different levels of substructure between voids. The MIP avoids general assumptions over void properties and provides bias measures for individual voids, allowing us to also see differences between voids.
The MIP simulation was also used to facilitate the study of the alignment of haloes in filaments and walls 
\citep{Aragon2014}. In this application the stacked density field was used to create one single detailed high-resolution LSS characterization. Using standard N-body techniques would have required either a large box with a large grid for LSS analysis or to perform the LSS analysis on a large number of smaller boxes for which the LSS would have to be characterized individually. Finally, we also took advantage of the high halo number density in the MIP simulation in the work presented in \citet{Wang13} to improve the velocity field analysis and to produce continuous maps of features in the velocity field. The same analysis on a single-simulation would have produced a sparse halo sample and a noisy velocity field. 

%==========================================================================
%--- Ensemble simulations
%==========================================================================
\section{Local Ensemble Simulations}\label{sec:realizations}

The overall structure and dynamics of the Cosmic Web is mainly defined by large-scale fluctuations, with smaller fluctuations
playing only a minor role \citep{Little91, Suhhonenko11, Einasto11}.
Since the primordial density field is a linear combination of random fluctuations it can be separated into two regimes
divided by a cut-off scale $S_{\textrm{\tiny{cut}}}$:
one regime being responsible for shaping the large-scale features of the Cosmic Web while the other the small-scale galaxy-size fluctuations as:

\begin{equation}\label{eq:separate_field} 
\delta = \delta(< k_{\textrm{\tiny{cut}}}) + \delta(\ge k_{\textrm{\tiny{cut}}})
\end{equation} 

\noindent where $\delta$ is the total primordial density fluctuations field, $k_{\textrm{\tiny{cut}}} = 1/S_{\textrm{\tiny{cut}}}$ marks the division between large-scale fluctuations:

\begin{equation}
\delta(< k_{\textrm{\tiny{cut}}}) = \delta(k) \; W(k).
\end{equation}

\noindent and small-scale fluctuations:

\begin{equation}
\delta(\ge k_{\textrm{\tiny{cut}}}) = \delta(k) \; (1-W(k)), 
\end{equation}

\noindent where $W(k)$ is a filter with a cutoff frequency $k_{\textrm{\tiny{cut}}}$ given by:

\begin{equation}\label{eq:k_filter}
W(k) = \qquad \left\{
 \begin{array}{rl}
     0, \qquad & k > k_{\textrm{\tiny{cut}}} \\
     1, \qquad & k \le k_{\textrm{\tiny{cut}}}. \\
 \end{array} \right.
\end{equation}

\noindent Following equation \ref{eq:separate_field}  we can generate multiple correlated realizations from the same 
primordial density field by fixing the large-scale fluctuations $\delta(< k_{\textrm{\tiny{cut}}})$ between realization and 
allowing the small-scale fluctuations $\delta(\ge k_{\textrm{\tiny{cut}}})$ to vary. 
The choice of $k_{\textrm{\tiny{cut}}}$ determines the scale and equivalent mass below which realizations are independent between each other:

\begin{equation}
M(S_{\textrm{\tiny{cut}}}) = \frac{4}{3}\pi S_{\textrm{\tiny{cut}}}^3 \bar{\rho}.
\end{equation}

\noindent Density fluctuations below this scale/mass are statistically independent in the linear regime and latter become correlated by the power transfer from large scales. 
Haloes with masses larger than $M(S_{\textrm{\tiny{cut}}})$ are correlated across the ensemble offering a unique opportunity to
study their persistent ensemble properties. These properties reflect the large-scale environment
of the halo as the small-scale contributions are averaged-out by the ensemble.

\subsection{Nested ensembles}

The idea of a correlated ensemble can be extended to iteratively generate nested correlated ensembles where for each realization $i$:

\begin{equation}
\delta^i = \delta(< k_{\textrm{\tiny{cut}}}) + \delta(\ge k_{\textrm{\tiny{cut}}})^i
\end{equation} 

\noindent we generate a new ensemble defined by a new cut-off mode $k_{\textrm{\tiny{cut2}}} > k_{\textrm{\tiny{cut}}}$. 
Realizations in the nested ensemble will share structures at scales larger than
$S_{\textrm{\tiny{cut2}}} = 1/k_{\textrm{\tiny{cut2}}}$ and will be independent below those scales:

\begin{equation}
\delta_2^i = \delta(< k_{\textrm{\tiny{cut}}}) + \delta(\ge k_{\textrm{\tiny{cut}}}, < k_{\textrm{\tiny{cut2}}})^i + \delta(\ge k_{\textrm{\tiny{cut2}}}).
\end{equation} 

\noindent Here the subscript denotes the level of recursion in the ensemble.

Designing a nested correlated ensemble one could define a cutoff mode $k_{\textrm{\tiny{cut}}}$ corresponding to cluster-group size structures 
and having a galaxy-size independent halo population between realizations. The next recursion level would be then determined by a higher $k_{\textrm{\tiny{cut2}}}$ corresponding to galaxy size haloes thus generating an independent population of sub-haloes between realizations for every galaxy-size halo in the first recursion level of the ensemble. 

The number of realizations grows with the number of nested ensembles as $N^m$ where $N$ is the number of realizations per ensemble
and $m$ is the level of recursion. This exponential behaviour makes it unfeasible to generate more 
than a just few recursion levels. In the present work we present a one-level ensemble. A two-level nested ensemble is in preparation 
at the time of writing.

%=======================================
%
%=======================================
\subsection{Practical advantages}

The self-contained nature of each realization in the correlated ensemble offers several practical advantages over standard large N-body simulations:

\begin{itemize}
\item Trivial to parallelize (run and post-processing). Realizations are self-contained and can
be run and analyzed independently.
\item Computing time increases linearly with the number of realizations.
\item Adding more realizations is trivial as it only requires generating new small-scale fluctuations.
\item No need for custom read-write routines as in the case of standard massive simulations where parallel
IO routines are necessary given the shear size of the datasets.
\item Efficient storage of many relatively small snapshots.
\item Its distributed nature makes it ideal for distributed architectures with efficient IO such as datacenters where it can be analyzed using Big Data tools \citep{Stickley15}.
\end{itemize}

It could seem that the above advantages are the same as if we simply ran a large number of independent realizations. 
However, in our case each realization corresponds to the same LSS configuration so all realization can be effectively
considered to be the same simulation with different halo formation paths.

%==========================================================================
%--- Computing local statistics
%==========================================================================
\section{Local Ensemble Statistics vs. Ergodicity}

The concept of ensemble implies the existence of statistically independent events. The ensemble average of a variable of
the system is given by 

\begin{equation} \label{eq:ensemble_average}
\langle w \rangle  =  \sum_{i=1}^N w_i p_i, 
\end{equation}

\noindent where $w_i$ is the measured variable at event $i$ and $p_i$ is the probability of observing the event $i$, with
$p_i = 1/N$ when all events have the same probability of being observed. The $n^{\textrm{\tiny{th}}}$ central moments
of a variable with respect to its ensemble average are:

\begin{equation}
\langle |w- \langle w \rangle |^n \rangle = \lim_{N \to \infty} \frac{1}{N} \sum_{i=1}^N (w_i - \langle w \rangle )^n .
\end{equation}

\noindent where $\langle w \rangle $ is the ensemble average defined in equation \ref{eq:ensemble_average} and $N$ is the number of realizations
or observations in the ensemble. In standard N-body simulations it is common to invoke ergodicity when computing halo properties and interchange volume 
averages by ensemble averages. In our case this is not necessary. Since realizations in the ensemble are correlated 
we can ``stack" them and instead define the \textit{local ensemble moments} of the stacked ensemble at position $\mathbf{x}$ as:

\begin{equation}
\langle |w(\mathbf{x}) - \langle w(\mathbf{x}) \rangle |^n \rangle = 
    \lim_{N \to \infty} \frac{1}{N} \sum_{i=1}^N (w(\mathbf{x})_i - \langle w(\mathbf{x}) \rangle )^n .
\end{equation}

\noindent In practice we compute the local $n^{\textrm{\tiny{th}}}$ central ensemble moment by considering all $M$ 
sampling points (haloes or particles) inside a sampling window of radius $R$ located at position $\mathbf{x}_i$ across all $N$ realizations:

\begin{equation}
\langle |w(\mathbf{x})- \langle w(\mathbf{x}) \rangle_R |^n \rangle_R = \frac{1}{N} \sum_{j=1}^N \frac{1}{M} \sum_{i=1}^M (w(\mathbf{x}_{i,j}) - \langle w(\mathbf{x}) \rangle_R )^n .
\end{equation}

\noindent where $w(\mathbf{x}_{i,j})$ is the variable measured at sampling point $\mathbf{x}_{i}$ inside the sphere corresponding to
realization $j$ and $\langle w(\mathbf{x}) \rangle_R $ is the local ensemble mean.

\subsection{Realization stacking and signal-to-noise}\label{signa_to_noise}

The stacking of realizations has the effect of ``averaging-out" random variations between realization. This is similar to the ``image stacking" technique
used in image processing for noise reduction. The signal-to-noise ratio (s$/$n) of the stacked ensemble depends on the number of realizations of the ensemble ($N$) as:

\begin{equation} \label{eq:signal_noise}
   \textrm{s$/$n} \propto \frac{N}{\sqrt{N}}.
\end{equation}

\noindent Here the signal-to-noise gives us an indication of the gain in the measured signal with respect to a single realization. In practice it is possible to generate enough realizations, using the technique described above, in order to increase the signal-to-noise ratio to any desired level.

%==========================================================================
%--- Simulation and haloe/subhalo analysis
%==========================================================================
\section{The MIP Simulation: Setting Up}

In this section we describe the \textit{Multum In Parvo}\footnote{Latin for ``much in little". The expression is 
also used in computer graphics to denote image pyramids (mipmaps) used in texture rendering.} (MIP) ensemble simulation. The MIP simulation
is an undergoing project consisting (in its first stage) of 220 realizations of a 32 $h^{-1}$Mpc box, each containing  $256^3$ particles, 
giving a mass per particle of $1.62\times10^{8}$M$_{\odot}$h$^{-1}$.
We adopted a $\Lambda$CDM cosmology with parameters $\Omega_m = 0.3$, $\Omega_{\Lambda}=0.7$, h=0.73,
$\sigma_8 = 0.84$ and spectral index $n=0.93$, close to values measured by the Planck mission \citep{Planck15}. The exact value of the cosmological parameters is not crucial for our purposes since we are interested on local variations with respect to the Cosmic Web,  which would likely be only modulated by small changes in the cosmological parameters.

%-------------------
%----  FIGURE  -----
%-------------------
\begin{figure}
  \centering
  %---
  \includegraphics[width=0.49\textwidth,angle=0.0]{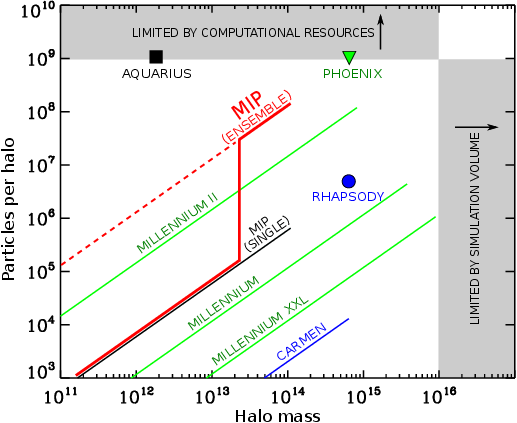}
    \caption{Comparison of the number of particles per haloes mass between the MIP and several simulations in the literature. 
	Lines indicate simulations with a ``continuous" halo population and
	isolated symbols indicate a resimulation of individual haloes. The single-realization MIP is shown as a black line. The MIP ensemble 
	simulation is shown in red. The lower regime of the MIP line corresponds haloes with masses below $M(k_{\textrm{\tiny{cut}}})$. In this mass regime
	individual haloes are independent between realizations and can not be stacked into an ``ensemble halo", thus the number of particles per halo is the same as in a 
	single realization. The
	upper regime of the MIP simulation corresponds to haloes with masses larger than $M(k_{\textrm{\tiny{cut}}})$. In this mass regime haloes
	can be stacked as they are cross-correlated across the ensemble and the number or particles per halo scales with the number of realizations.
	The MIP simulation occupies a unique region in the diagram in terms of its two-regime behaviour and unprecedented particle
	count for a continuous halo population. It has the highest number of particles per halo above 
	M$_{\textrm{\tiny{cut}}}$.} 
  \label{fig:mip_comparison}
\end{figure}

The results presented here correspond to 220 realizations that have been completely run and analyzed to identify FoF haloes and SubFind subhaloes 
in all snapshots. The full MIP ensemble simulation is equivalent in terms of volume and number of particles to a 256 $h^{-1}$Mpc box standard
simulation with $2048^3$ particles. The current status of 220 realization is equivalent to a $32 \times (220)^{1/3} \sim 193$ h$^{-1}$Mpc box with
$256 \times (220)^{1/3}\sim 1540^3$ particles. 
%A comparison between the properties of a single realization and the ensemble simulation is presented in Table \ref{tab:simulations}.

%-------------------
%----  FIGURE  -----
%-------------------
\begin{figure*}
  \centering
  %--- 
  \includegraphics[width=0.98\textwidth,angle=0.0]{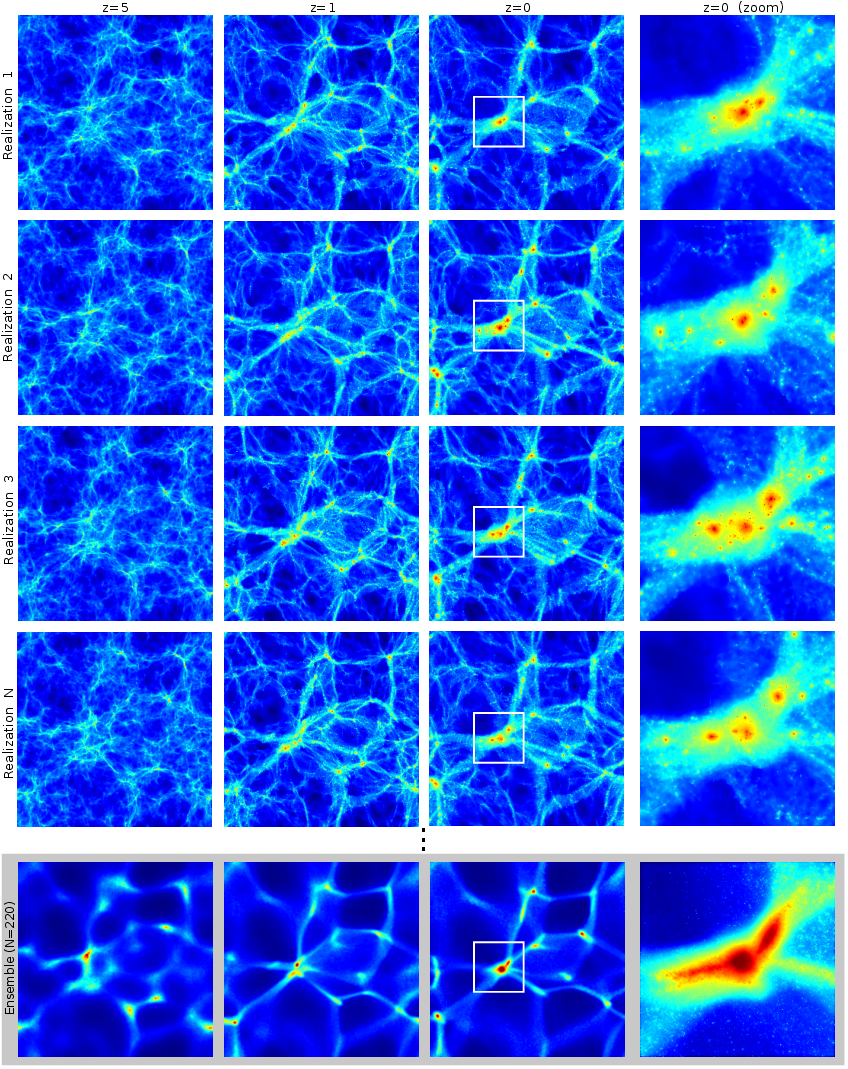}
    \caption{{\bf Comparison between four realizations of the MIP ensemble}. Colors show the logarithmically scaled dark matter density field (going from dark blue to dark red with increasing density) on a 2 h$^{-1}$Mpc thick slice across the simulation box for 
	four realizations (top four rows) and the stack of 220 realizations (bottom row) at
    redshifts $z=5,1$ and $0$ (first three columns from left to right). A zoom region of $8 \times 8$ h$^{-1}$Mpc centred on a large group that is present in all realizations of the ensemble is shown on the right column. All the realizations in the ensemble contain the same LSS features but a unique halo population. This is clearly seen in the zoom regions where the same group has a different internal structure in different between realizations.}
  \label{fig:ensemble_simulation_grid}
\end{figure*}

%=======================================
%
%=======================================
\subsection{Generating initial conditions}

The first step in making the correlated ensemble is the creation of an initial conditions ``template" from which all realizations in the ensemble will be spawned. The template field was generated using the publicly available 
{\small GRAFIC2} code \citep{Bertschinger01}.
From the template each of the semi-independent realizations was generated as described in section \ref{sec:realizations}. 
In practice we do not generate new small-scale fluctuations for each realization but instead shift 
$\delta(< S_{\textrm{\tiny{cut}}})$ in real space along each dimension by a multiple of
$S_{\textrm{\tiny{cut}}}$ which in our case corresponds to 4 h$^{-1}$Mpc. A ``cell" of 4 h$^{-1}$Mpc of side contains a mass
of $\sim 2 \times 10^{13}$ h$^{-1}$ M$_{\odot}$. 
For the 32 h$^{-1}$ Mpc box this gives $32/4=8$ independent shifts per dimension for a total of $8 \times 8 \times 8 = 512$ realizations (note that only 220 realizations were run).
From the set of initial condition files we then generated Gadget files starting at $z \simeq 80$ following the standard 
Zel'dovich prescription \citep{Zeldovich70}.

%=======================================
%
%=======================================
\subsubsection{Ensemble run}

Each realization was run using 16 processors on the Homewood High Performance Cluster (HHPC) at JHU. We stored 50 snapshots separated in logarithmic intervals in scale 
factor starting at $z=5$. The realizations were run and processed in batches of 20 runs in order to keep a low load on the HHPC cluster. 
All the snapshots from the 220 realizations occupy approximately 5.5 TB. Each ``ensemble snapshot" occupies
$\sim 100$ GB in contrast to the much smaller ``single-realization snapshot" with only $\sim 460$ MB.

%================
%--- Table
%================
%\input{table_simulations}\label{tab:emsemble_properties}

%=======================================
%
%=======================================
\subsubsection{MIP compared to standard simulations}

Figure \ref{fig:mip_comparison} shows a comparison between the MIP and several other simulations described in the literature:
Carmen \citep{McBride09}, Millennium \citep{Springel05}, Millennium II \citep{Boylan09} and Millennium XXL \citep{Angulo12}. We also show three high-resolution zoom resimulations: 
Aquarius \citep{Springel08}, Phoenix \citep{Gao12} and Rhapsody \citep{Wu12}. This is by no means a complete
list. A single realization from the MIP is, for current standards, a small-box medium-resolution simulation. It sits between the 
Millennium and Millennium II simulations in terms of mass resolution but it contains a much smaller volume. On the other hand, 
the MIP simulation has the highest number of particles per halo in its ``stacked ensemble" mode. 
Massive haloes (M $>$ M$ _{\textrm{\tiny{cut}}}$) are persistent across the ensemble and show small variations between realizations allowing us to study their 
ensemble statistical properties.
Haloes less massive than M$_{\textrm{\tiny{cut}}}$ are independent between realizations and can be stacked to increase the local 
halo number density. The MIP simulation has a halo number density 220 times higher than an equivalent single-realization simulation. For comparison, the density of \textit{haloes} 
in the MIP simulation is almost two times higher than the density of \textit{particles} in the Millennium I simulation!
Given its unique properties, one can not directly compare the MIP ensemble simulation to other standard simulations. The MIP
simulations stands apart from standard simulations in terms of particles per halo, halo number density, statistical properties,
and computational requirements.

%=======================================
%
%=======================================
\subsection{Haloe/subhalo catalogues} \label{sec:halos_subhalos}

We identified Friends of Friends (FoF) haloes for all 50 snapshots of each realization using a linking parameter of $b=0.2$.
Only haloes with a minimum mass of $3.2 \times 10^{9}$ M$_{\odot}$, corresponding to 20 particles, were included in the catalogue. 
For each halo we computed physical properties such as mass, radius, inertia tensor, angular momentum and $v_{\textrm{\tiny{max}}}$.
Subhaloes were identified using the SubFind code \citep{Springel01} which identifies subhaloes as 
gravitationally bounded substructures inside FoF haloes. We also computed physical properties for the subhaloes 
and stored both halo and subhalo catalogues in a database for efficient retrieval.

Given the large number of realizations, the creation, running and analysis of each of the 220 realizations is controlled
with an automated pipeline. The final output is a set of haloe/subhalo catalogues and their most massive progenitor 
lines.

%-------------------
%----  FIGURE  -----
%-------------------
\begin{figure*}
  \centering
  \includegraphics[width=0.99\textwidth,angle=0.0]{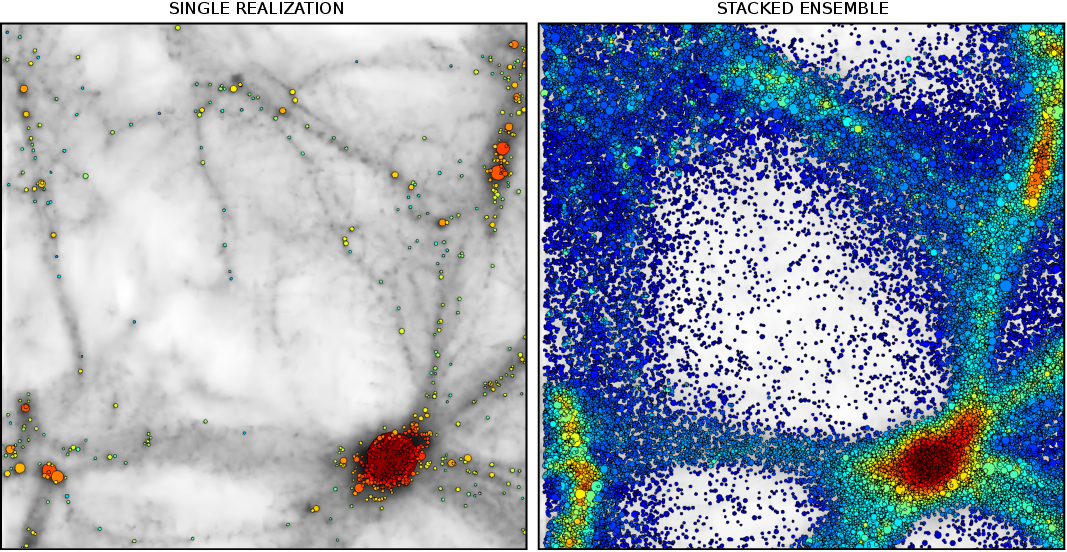}
    \caption{{\bf Single realization vs. ensemble stack}. 
     Circles correspond to haloes identified inside a 2 h$^{-1}$ Mpc thick slice across the simulation box. The circles are scaled with the
	haloe's radius and colored with the value of the ensemble-averaged density field at the haloe's position. 
	For clarity we only show subhaloes with radius smaller than 500 h$^{-1}$ kpc.
	The gray background corresponds to the density field for one single realization (left) and the ensemble (right). The halo number density of the stack is significantly higher compared
	to the single realization. Even the most underdense regions inside the large void are populated by haloes.}
  \label{fig:single_vs_ensemble_void}
\end{figure*} 

%-------------------
%----  FIGURE  -----
%-------------------
\begin{figure*}
  \centering
  %---
  \includegraphics[width=0.99\textwidth,angle=0.0]{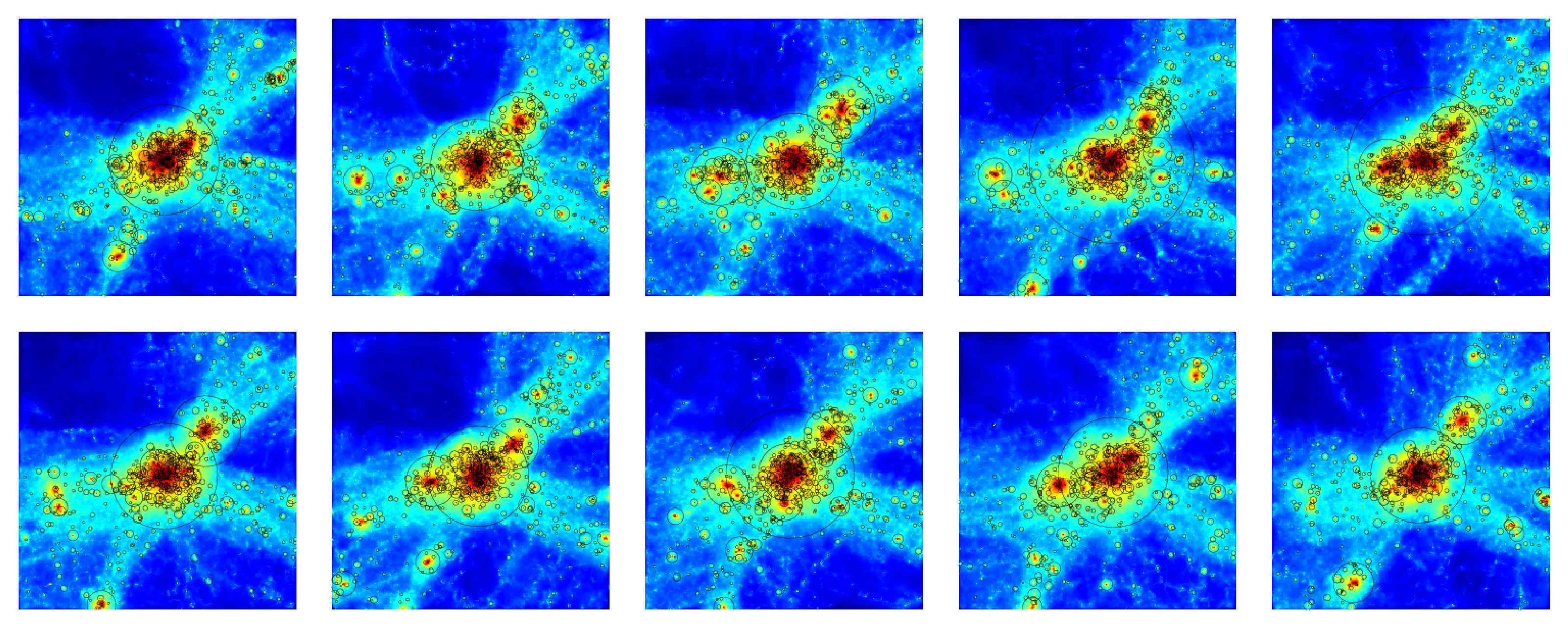}
    \caption{Ten realizations of the central cluster shown in Fig. \ref{fig:ensemble_simulation_grid}. The SubFind
	subhaloes are shown as open circles scaled with their radius. The background image represents 
	the dark matter density field of the corresponding realization.}
	\label{fig:thirty_clusters}
\end{figure*}

%=======================================
%
%=======================================
\section{The MIP Simulation: First Results}

We now describe the general properties of the MIP simulation starting with individual realizations, the
global mass function (per realization and ensemble) and the basic properties of the most massive haloes in the simulation.

\subsection{Individual realizations}
 
Figure \ref{fig:ensemble_simulation_grid} shows an overall view of the MIP ensemble simulation. Four individual realizations are shown as well
as the stacked ensemble at redshifts 
$z=5,1$ and $0$ and a zoom into the central halo in the slice. 
Already at $z=5$ we can identify common structures between realizations like 
the large void at the top-left corner and the overdense region that will collapse to form the largest halo at the centre of the slice. 
On the other hand, the small-scale fluctuations are clearly different between realizations.
This becomes more evident as the simulation evolves and the fluctuations collapse into haloes with 
unique substructure. 
At $z=1$ it is clear that even when the LSS is roughly the same in all realizations, it is composed by a unique halo population. 
This can be seen in all environments but it is more significant in walls and voids where the halo number density is low 
and often the same region of space is unevenly sampled between realizations. All realizations produce a similar cluster at the centre of the slice in Fig. \ref{fig:ensemble_simulation_grid}. However, a detailed view inside the cluster shows 
significant differences at small scales. For instance, the central cluster in the first two realizations 
of Fig. \ref{fig:ensemble_simulation_grid} has two large subhaloes while the last two realizations have three large subhaloes. 
The differences at smaller scales are even larger making it impossible to identify common low-mass subhaloes between realizations. 
The halo population corresponding to these scales is effectively independent between realizations.

%---------------------------------
%
%---------------------------------
\subsection{Stacked ensemble}

The stacked density field of the ensemble is shown in the bottom panels of Figure \ref{fig:ensemble_simulation_grid}. 
It has some interesting properties that make it useful for applications where a smooth field is needed.
The stacking of realizations acts similar to a low-pass filter producing a smooth density field, while at the same time retaining anisotropic 
features, in contrast to what we would expect if the field was smoothed with a Gaussian kernel. Note that the stacked density field 
looks remarkably similar in the three redshifts shown.
The ensemble density field can be used as a robust large-scale density field tracer which is not affected by
small scale fluctuations such as haloes, which tend to negatively affect structure finder techniques based
on the density field. A good illustration of this is the thin filament located at the bottom-right of the central cluster. 
In all four realizations there are significant differences in the substructures defining the filament. The ensemble 
density field, on the other hand, shows a well defined tenuous structure  with practically no substructure and 
an almost constant density profile. The ensemble density field looks similar to a Lagrangian-smoothed density field 
see \citep[see][]{Little91, Melott93, Avila01, Suhhonenko11, Aragon10b, Aragon12a} where the primordial density field is 
smoothed in the linear regime and then evolved. Such Lagrangian-smoothed simulations contain only large-scale modes and 
no substructure below the smoothing scales and therefore can not be directly used to study halo populations. Figure \ref{fig:single_vs_ensemble_void} shows a comparison between the haloes in a single realization and the stacked ensemble in a region containing a cosmological void. While the single realization contains a handful of haloes inside the void, preventing any statistical analysis, the stacked ensemble has a high number density of haloes even in the most underdense regions of the void.

%-------------------
%----  FIGURE  -----
%-------------------
\begin{figure}
  \centering
  \includegraphics[width=0.48\textwidth,angle=0.0]{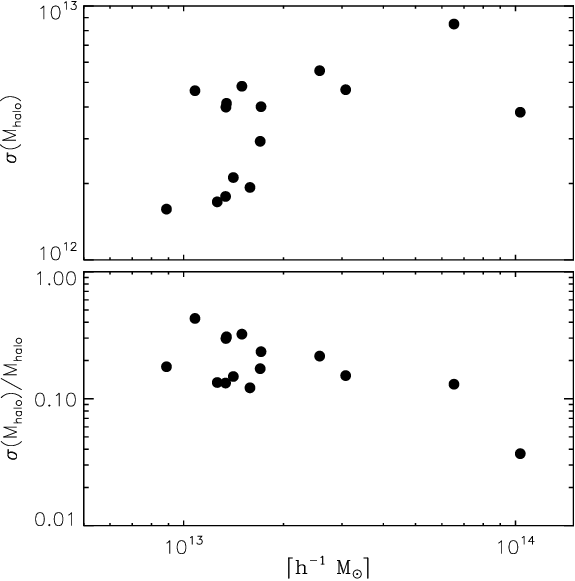}
    \caption{Top: halo mass dispersion across the ensemble for the 15 most massive haloes in the simulation. Bottom: relative
	halo mass dispersion.}
  \label{fig:halo_mass_dispersion}
\end{figure} 

%-------------------
%----  FIGURE  -----
%-------------------
\begin{figure}
  \centering
  \includegraphics[width=0.48\textwidth,angle=0.0]{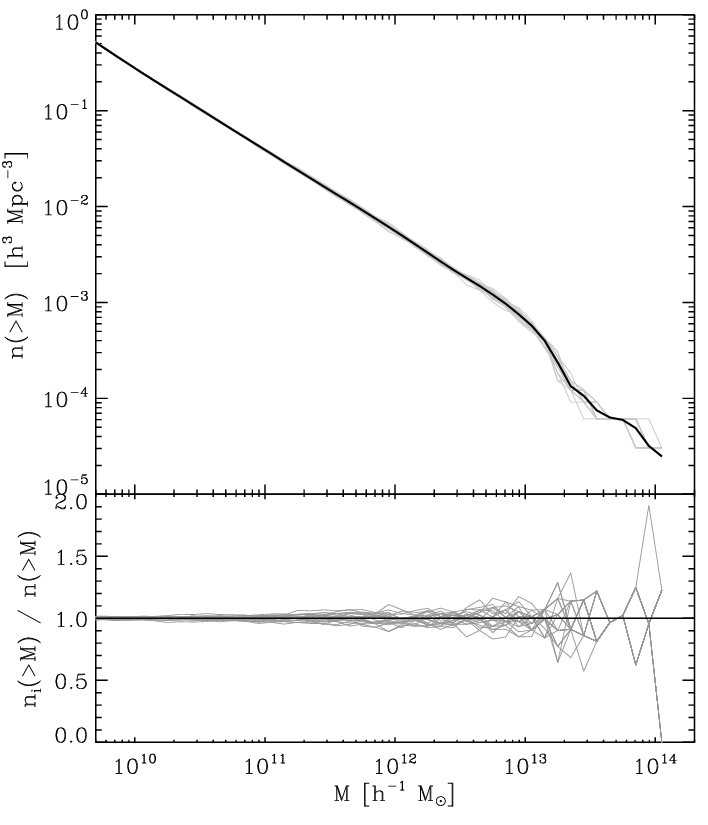}
    \caption{{\bf Global halo mass function}. Top: the black solid line corresponds to the halo mass function of the entire ensemble. 
		The thin gray lines show the halo mass function computed from 20 individual realizations.
		Bottom: ratio between the mass function of each realization and the ensemble mass function.}
  \label{fig:ensemble_mass_function}
\end{figure} 

%=======================================
%
%=======================================
\subsection{Ensemble halo population}

Figure \ref{fig:thirty_clusters} depicts 10 versions of the central cluster shown in 
Figure \ref {fig:ensemble_simulation_grid} identified in different realizations. 
The ensemble mean mass of the central FoF halo is $\sim 10^{14}$ h$^{-1}$M$_{\odot}$.
We highlight the substructure inside the cluster by the position of the subhaloes as open circles scaled with their radius.
While the megaparsec-scale  matter distribution is roughly similar in all realizations, there is a large
dispersion in the properties of the haloe/subhalo populations. This can be seen in the most massive
subhalo of the cluster which has a difference in its size of almost a factor of two between the most extreme 
versions of the cluster across the ensemble. Each realization of the cluster shows a unique formation path with 
different halo mass, halo accretion histories, internal dynamics etc. However, because in all realizations the same
LSS is shared one expect the properties of the cluster to be highly correlated across the ensemble.

\subsubsection{Ensemble halo mass dispersion}

In order to get a better understanding of the differences between halo populations across the ensemble we computed 
the ensemble halo mass dispersion for the  15 most massive haloes in the simulation. 
We do this by first defining a  sample of reference haloes identified from a single realization.
Next we track the reference haloes across all realizations in the ensemble.
We track a given halo by placing a search window at the reference haloe's position in all realizations and selecting the most massive 
halo inside the search window. The search window is made equal to the reference haloes's virial radius.
Figure \ref{fig:halo_mass_dispersion} shows the ensemble mass dispersion for the 15 most massive haloes. The relative mass dispersion increases for decreasing halo mass,
confirming that haloes less massive than $\sim 10^{13}$ h$^{-1}$M$_{\odot}$ are defined by 
$\delta(\ge k_{\textrm{\tiny{cut}}})$.
The relative halo mass dispersion can be use as a guide for setting a threshold above which haloes can be 
studied on a case-by-case basis ( M$(k_{\textrm{\tiny{cut}}}) \sim 2 \times 10^{13}$ h$^{-1}$M$_{\odot}$ for the MIP). 

\subsubsection{Ensemble halo mass function}

Figure \ref{fig:ensemble_mass_function} shows the mass function computed from 20 individual realizations 
and the full ensemble. The mass function of individual realizations closely follows the ensemble mass function with some variations becoming
more significant for halo masses larger than $\sim 10^{13}$ h$^{-1}$ M$_{\odot}$, at which point the mass function becomes noisy 
due to the relatively small volume of the simulation box. The ``knee" in the mass function at M$\sim 10^{13}$ h$^{-1}$ M$_{\odot}$ is a 
particular characteristic of the initial conditions template used to create the ensemble. 
From the volume-mass defined by $S_{\textrm{\tiny{cut}}}$ we expect to have roughly the same
halo population above a mass of $\sim 10^{13}$ h$^{-1}$ Mpc.

%==========================================================
%
%==========================================================
\section{Illustrative application: Local variations in halo-LSS correlations}\label{sec:application_2}

In this section we present an exploratory study of the alignment of haloes in the Cosmic Web. We take advantage of the high number density of haloes in the MIP stacked ensemble to compute three \textit{local halo ensemble statistics}: halo ellipticity, halo shape-LSS alignment and the Pearson correlation between these two variables. Our goal is to show that even though there are global trends of these variables with cosmic environment  there can be large local variations even across a single cosmic structure, a feature not seen in standard haloe-LSS analysis. 

The haloe's ellipticity was determined on the basis of the ratio of the haloe's main axis as:

\begin{equation} 
e = 1 - \frac{c}{a},
\end{equation}

\noindent where $a = \sqrt{\lambda_1}$ and $c = \sqrt{\lambda_3}$ are the major and minor axis of the halo and $\lambda_1, \lambda_3$ are smallest and largest eigenvalues respectively of the inertia tensor:

\begin{equation}
\mathbf{I}_{i,j} = \frac{\sum_{i=1}^N m_i \; x_i \; x_j}{\sum_{i=1}^N m_i},
\end{equation}

\noindent where $m_i$ is the particle's mass and the positions $x_i, \; x_j$ are with respect to the centre of mass:

\begin{equation}
\bar{x} = \frac{1}{N} \sum_i^{N} x_i, 
\end{equation}

\noindent and the summation is over all particles in the halo ($N$).

%-------------------
%----  FIGURE  -----
%-------------------
\begin{figure}
  \centering
  \includegraphics[width=0.45\textwidth,angle=0.0]{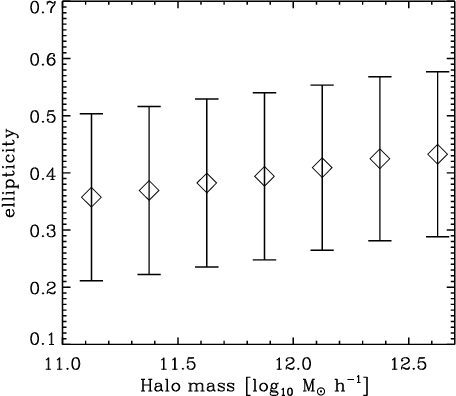}
  \caption{{\bf Ellipticity as a function of halo mass} The mean ellipticity inside each mass bin is indicated by the diamond symbol. The error bars indicate the dispersion of the ellipticity inside the bin.}
  \label{fig:global_ellipticity} 
\end{figure} 

The local LSS direction was computed using the MMF2 method described in \citep{Aragon2014}. The method is an extension of the original MMF \citep{Aragon07b} based on the local variations of the density field encoded in the Hessian matrix. Local directions of filaments are assigned as the smallest eigenvalue of the Hessian matrix \citep[see][]{Aragon07a}, computed at an equivalent scale of 2 h$^{-1}$Mpc.  The alignment between the halo's main axis of inertia and its host LSS, expressed as $\cos \theta$, can have values in the range [0-1] and has a mean value of $\cos \theta=0.5$ in the case of random orientation. Values higher than $\cos \theta=0.5$ indicate a parallel alignment while lower values a perpendicular alignment. Note that the filament orientation used here is dependent of the scale at which the Hessian is computed. Filaments tend to form inside larger sheet-like structures surrounding voids and as one moves to larger scales the smallest eigenvalue's direction will change from being located along the direction of the filament to an indetermined direction along the plane of its parent sheet.

\begin{figure*}
  \centering
  \includegraphics[width=0.99\textwidth,angle=0.0]{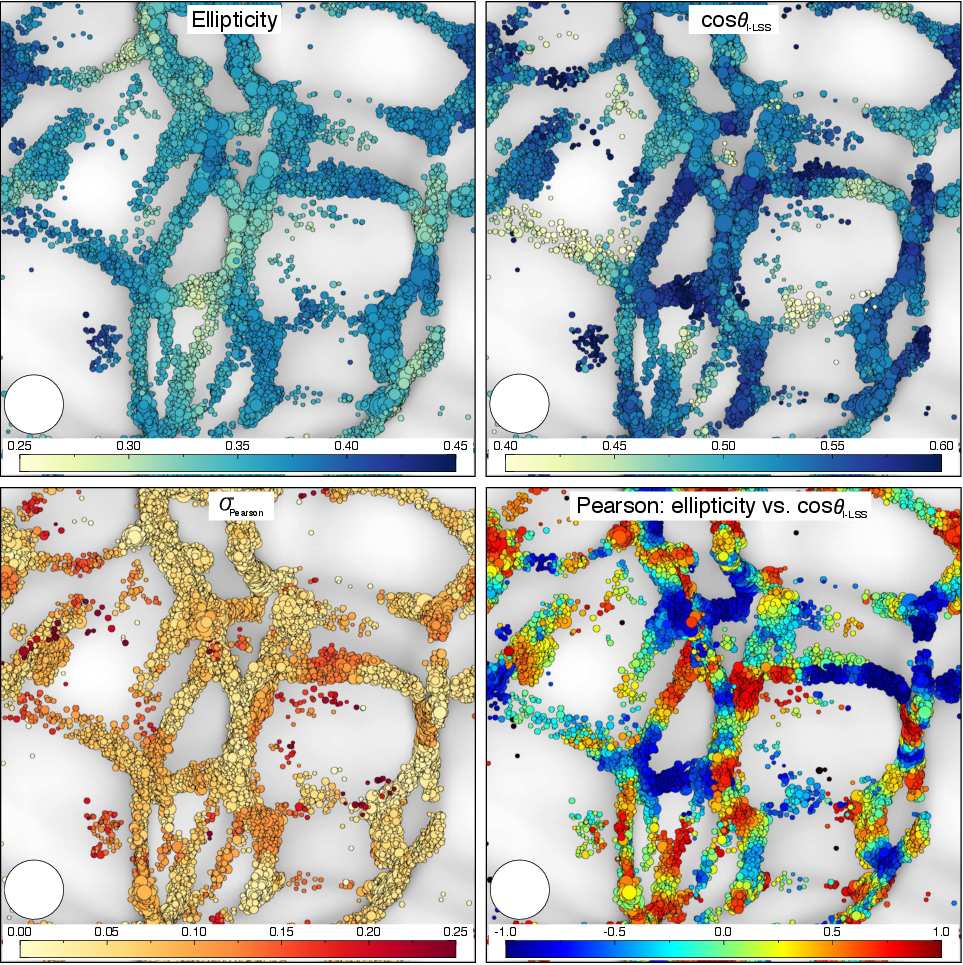}
  \caption{{\bf Local ensemble halo statistics} From top-left clockwise we show the values of the ensemble-average mean halo ellipticity, the halo shape-LSS alignment, the Pearson correlation between the halo ellipticity and halo shape-LSS alignment and the standard deviation of the Pearson correlation between the halo ellipticity and halo shape-LSS alignment ($\sigma_{\textrm{\tiny{Pearson}}}$). The colored circles correspond to haloes (scaled with their virial radius) located in filaments. The large white circles indicate the size of the sampling sphere used to compute the local ensemble statistics. The values corresponding to colours are indicated in the colour bars at the bottom of each panel. Note that the colour scale is different for each panel.}
  \label{fig:halo_props} 
\end{figure*}

Figure \ref{fig:global_ellipticity} shows the mean halo ellipticity as a function of halo mass computed from the whole simulation box. From this plot we can see a global trend in increasing ellipticity with increasing halo mass (although with a large dispersion).  While this analysis provides a quantitative global measurement of halo ellipticity, it is not able to identify possible variations in the ellipticity in different cosmic environments.  Figure \ref{fig:halo_props}, on the other hand, shows the local halo ellipticity computed from the stacked ensemble. Figure \ref{fig:halo_props} not only gives a visual impression of the mean ellipticity and even its dispersion just like Fig. \ref{fig:global_ellipticity} but more importantly, it shows the local variations of ellipticity in different regions of the Cosmic Web. halo ellipticity not only changes locally but is strongly correlated with local LSS. There are clear domains containing haloes with similar ellipticity. Note that this effect is enhanced by the averaging window which erases variations in the local ellipticity at scales smaller than the window. Nevertheless the size of the domains is in many cases larger than the averaging window and we can observe that filaments larger than the averaging window tend to have haloes with similar ellipticity. 

%-------------------
%----  FIGURE  -----
%-------------------

The information presented in Fig. \ref{fig:halo_props}, where we use colours to encode halo properties, may seem qualitative compared to a standard plot like Fig. \ref{fig:global_ellipticity}. However, Fig.  \ref{fig:halo_props} allows us to visually identify spatial trends given by the local environment and potentially unveil unknown environmental effects. 

One example of this is shown in the top-right panel of Fig. \ref{fig:halo_props} where we show the alignment between the haloe's main axis of inertia and the direction of it local LSS. In general, the main axis of inertia of a halo is oriented along the direction of its parent filament. However, there are important variations in the shape-LSS alignment between different regions. Note how the shape-LSS alignment has larger spatial variations compared to the more smooth spatial distribution of halo ellipticity. In particular, the filament near the centre of the plot, with an angle of $\sim 45$ degrees, shows a surprising behaviour. Haloes at both ends of the filament are more strongly aligned with the filament than haloes near the centre of the same filament!. This unexpected behaviour could not have been observed by a global analysis like the one presented in Fig. \ref{fig:global_ellipticity}. However, one could have seen this effect by measuring variations along individual filaments but we would have to anticipate such effects in the first place. Instead, Fig. \ref{fig:halo_props} allows us to study, without any assumptions, local effects in halo properties and discover new trends. The power of meaningful visualization of complex data can not be underestimated.

The ellipticity and shape-LSS orientation of haloes can be used to constrain models of galaxy formation as they reflect the anisotropic accretion of matter into haloes. The measurement of halo ellipticity,  halo shape-LSS and halo spin-LSS alignment from the observed luminous galaxies is a challenging task as the many, and often contradicting, works in the literature show \citep[][among others]{Trujillo06,Slosar09,Jones10,Lee11,Varela12,Zhang13}. In the case of elliptical galaxies one can assume that the distribution of light traces the projected shape of the galaxy's dark matter halo. From this we can derive the (projected) halo ellipticity, halo shape-LSS alignment and even halo spin-LSS alignment assuming that the angular momentum of a halo is aligned with its smallest axis \citep[][and references therein]{Allgood06}. However, as discussed in \citet{Tempel2014} the measurement of galaxy shapes tends to avoid spherical galaxies and favor more elliptical shapes, further complicating the reliable identification of ellipticity and shape.

In order to measure the degree of correlation between halo ellipticity and shape-LSS alignment we computed the (local)
Pearson correlation which quantifies linear correlation between two variables:

\begin{equation}
r = \frac{\sum_{i=1}^N (x_i - \bar{x})(y_i-\bar{y})}{ \sqrt{\sum_{i=1}^N(x_i-\bar{x})^2}  \sqrt{\sum_{i=1}^N(y_i-\bar{y})^2} }.
\end{equation}

\noindent Correlation and anti-correlations are indicated by values of 1 and -1 respectively, values in-between indicate a weaker correlation.
Figure \ref{fig:halo_props} shows the Pearson correlation between the halo ellipticity and shape-LSS alignment. Visual inspection indicates that ellipticity correlates well with shape alignment for most haloes and in most environments. This is encouraging for observations using elliptical shapes to derive halo shape and spin-LSS alignment.  However, the figure also shows regions where there is a significant anti-correlation between the two variables. One extreme case is the vertical filament at the centre of the figure. Its upper half shows strong correlation while its lower half strong anti-correlation! An exploration of the surrounding LSS shows a large group close to the anti-correlation region. This figure suggest that in order to obtain a reliable spin alignment signal when using galaxy shapes one should avoid galaxies near large groups or clusters. The effect is likely the result of the complex tidal field in those regions. Indeed, the study of \citet{Jones10} shows that, in the case of spin-LSS alignment,  the signal is stronger in regions far from strong gravitational perturbations. Figure \ref{fig:halo_props} also shows the standard deviation of the Pearson correlation signal, computed by generating 100 samplings of the ensemble with (randomly selected) half the number of haloes and then computing the standard deviation of the samplings. Despite there being large spatial variations in the Pearson correlation its standard deviation is relatively small and has also small spatial variations. The standard deviation across the ensemble seems to be larger in low density regions and it is low ($\sigma_{\textrm{\tiny{Pearson}}} < 0.1$) in the filament highligted in the discussion above. 

By computing local statistics instead of global we are able to directly observe the complex correlations between halo properties and cosmic environment and infer relations that otherwise would have remained unknown.  The application of diagrams such as Fig. \ref{fig:halo_props} (and the technique presented here) will prove useful in the study of individual cosmic structures such as the local supercluster where we are not only interested in global properties but on the local behaviour of the galaxies in our particular cosmic environment.

In order provide an estimate of the level of contamination in halo shape-based studies arising from changes in the orientation of haloes along a given structure we performed a simple quantitative analysis of the number of filaments in which the correlation between ellipticity and shape-LSS changes along the same filament. Using the filaments identified with the MMF2 method \citep{Aragon2014} we divided the box into slices of 2 Mpc thickness and proceeded to count the number of filaments in the slice with significant changes in their ellipticity vs. shape-LSS alignment. If the Pearson correlation value between the two extremes of the filament was larger than $\Delta = 1$ then the filament was labeled as a \textit{transition filament}. In addition to that we also required that the difference in the correlation was centred on zero. We found the fraction of transition filaments to be $\sim 0.2$ compared to the filaments with haloes showing a similar correlation signal across their length. Transition filaments are mostly limited to filaments connected to massive clusters. Although our simulation box is not large enough to contain a large sample of massive clusters it gives us a clear indication of the effect of clusters in the shape-LSS alignment signal. Studies focused of halo-LSS alignment based on observed elliptical galaxies can be affected by contamination when including galaxies in filaments connected to massive clusters. This may be one of the reasons, appart from the intrinsic complexity of LSS analysis, for the large discrepancies between reported alignment signals in early (mostly cluster-based) studies of halo-LSS alignment.

%==========================================================================
%--- Conclusions
%==========================================================================
\section{Conclusions and Future Prospects}

We introduced a novel application of correlated ensembles in which realizations share the same large-scale fluctuations while having independent small-scale fluctuations. This technique represents a significant improvement over standard single-realization N-body simulations. By generating a large set of semi-independent realizations and stacking them we break the fundamental limit in halo number density imposed by the halo mass function. This allows us to trace the distribution of haloes in all cosmic environments with unprecedented detail, from the dense clusters down to the under-dense voids. 

We presented the \textit{Multum In Parvo} (MIP) simulation/project consisting of a correlated ensemble simulation with 220 realizations. The MIP simulation is equivalent in terms of volume and number of particles to a 193 h$^{-1}$ Mpc box containing $\sim 1540^3$ particles and $\sim 5\times 10^6$  haloes with a minimum halo mass  of $3.25 \times 10^9$ h$^{-1}$ M$_{\odot}$. 

The unprecedented halo number density of the MIP simulation allows us to compute ensemble statistics on a local basis.  We presented an exploratory analysis of the local variations in the halo ellipticity and halo shape-LSS alignment as well as their correlation. While there are clear global trends, as reported in the literature, there are significant variations between individual cosmic structures. In particular we found that the correlation between halo ellipticity and halo shape-LSS orientation is affected when the filament is connected to a massive clusters. 

Given its broad scope, the technique presented here can be applied to a variety of physical phenomena to study the effect of environment on haloes and even cosmic structures given a proper choice of $k_{\textrm{\tiny{cut}}}$. Future applications of the MIP technique include the study of the ensemble statistics of the local Universe using constrained realizations and a more complete and detailed study of local variations of halo properties in order to identify cosmic environments where such properties can be reliably measured in the real Universe. 

%==========================================================================
%--- Acknowledgements
%==========================================================================
\section{Acknowledgements}
This research was partially funded by the Gordon and Betty Moore foundation. The author would like to thank Mark Neyrinck, Bernard Jones and Joe Silk for many stimulating discussions and two anonymous referees for their useful comments and suggestions.

%==========================================================================
%--- References
%==========================================================================

\end{document}